\documentclass{JAIS}

\journal{JAIS-ID}
\vol{2021}

\received{xx January 2018}
\published{xx March 2018}

\def\be{\begin{equation}}
\def\ee{\end{equation}}
\def\bea{\begin{eqnarray}}
\def\eea{\end{eqnarray}}

\usepackage{caption}

\begin{document}

\title{A Novel Readout Scheme for Muon Tomography Application in Material Identification}

\author{Subhendu Das,\auno{1,2} Sridhar Tripathy,\auno{1,2} Sandip Sarkar,\auno{1,2}, Nayana Majumdar\auno{1,2} and Supratik Mukhopadhyay\auno{1,2}}

\address{$^1$Saha Institute of Nuclear Physics, Kolkata, INDIA}
\address{$^2$Homi Bhabha National Institute, Mumbai, INDIA}

\begin{abstract}

This work reports a cost-effective, multi-parameter readout and data-acquisition system for a muon scattering tomography system based on Resistive Plate Chambers (RPCs). Initial test measurements with a prototype Resistive Plate Chamber were performed using a low-cost FPGA coupled to the NINO ASIC for the event selection and handling data. The Time over-Threshold (TOT) property of NINO ASICs has been used to achieve better position information and achieve precise tracking capability. In our test setup, we try to build an imaging setup using a single RPC and lead block, which covers some areas of RPC. A glass RPC of dimension 30cm x 30cm, filled with a gas mixture of 95\% Freon and 5\% Isobutane, equipped with two orthogonal panels of readout strips of width 3 cm and pitch 3.2 cm, has been operated.

\end{abstract}

\maketitle


\section{Introduction}
\label{sec:Introduction}
Muon tomography is a system for material identification utilizing multiple Coulomb scattering suffered by cosmic ray muons while passing through a matter. The resultant deflection from the original trajectory can be represented by a Gaussian distribution dependent on several physical properties of the matter and also the muon momentum. Thus the measurement of the scattering angle by tracking the pre and post-interaction muon trajectories enables one to identify the material.

\begin{figure}[h!]
\centering
\includegraphics[width=0.4\textwidth]{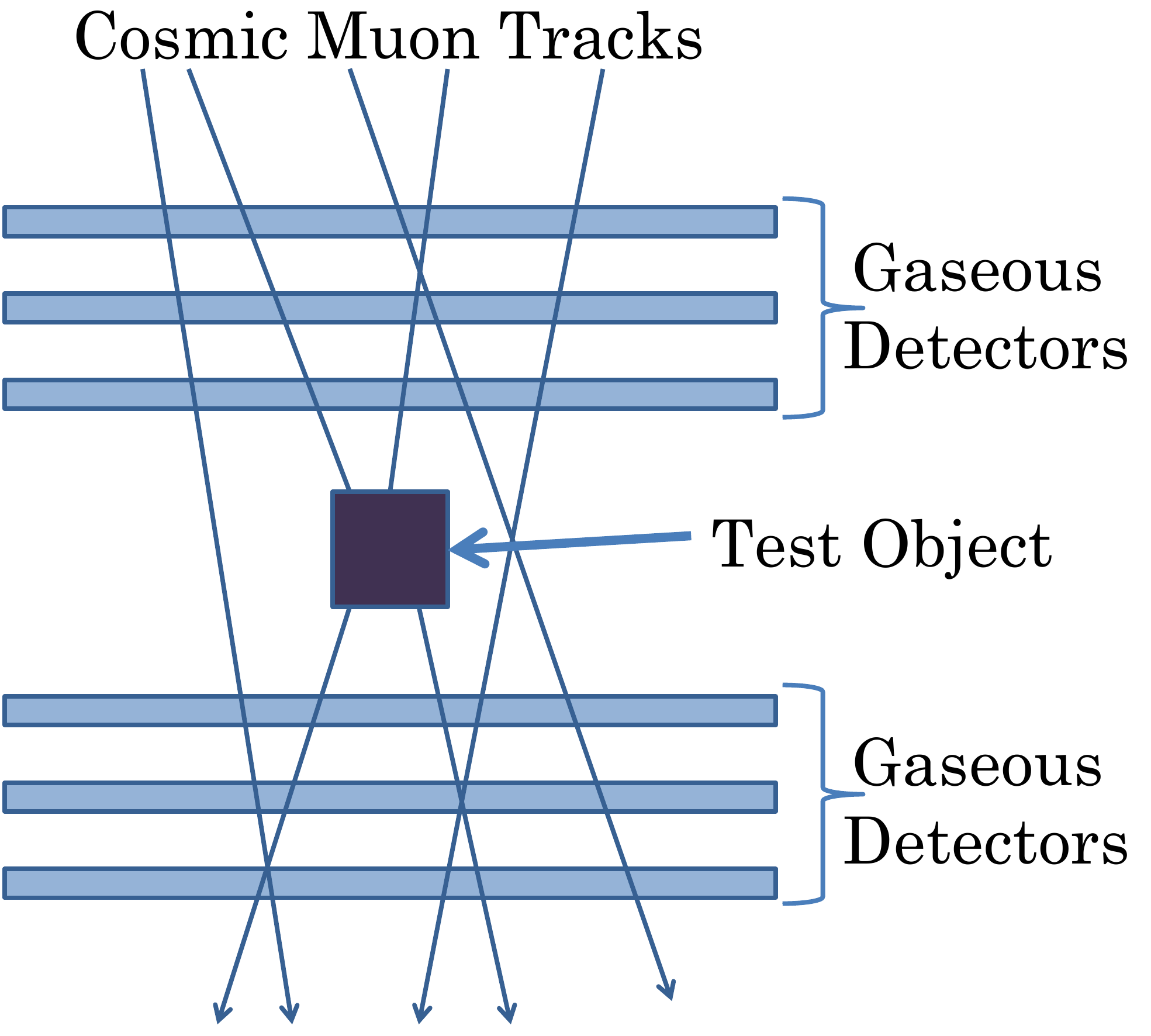}
\caption{Schamatic diagram of muon tomography system}
\label{fig:tomography system}
\end{figure}

We plan to use Resistive Plate Chambers (RPCs) for tracking muons before and after their interaction. The Resistive Plate Chamber
(RPC) is one of the most popular gaseous detectors which is suitable for this purpose due to its simple design, ease of construction, cost-effectiveness, along very good temporal, spatial resolutions, and detection efficiency. The spatial resolution of the RPC determines the precision in tracking and hence the quality of image formation.

An FPGA-based multi-parameter data acquisition system has been developed in this context for the collection of position information and subsequent track reconstruction. FPGA-based DAQ system is a simple low-cost multi-parameter scalable data acquisition system. This is useful for any kind of particle tracking experiment and imaging application.

\section{Readout Scheme}
\label{sec:Readout Scheme}
\begin{figure}[h!]
\centering
\includegraphics[width=0.9\textwidth]{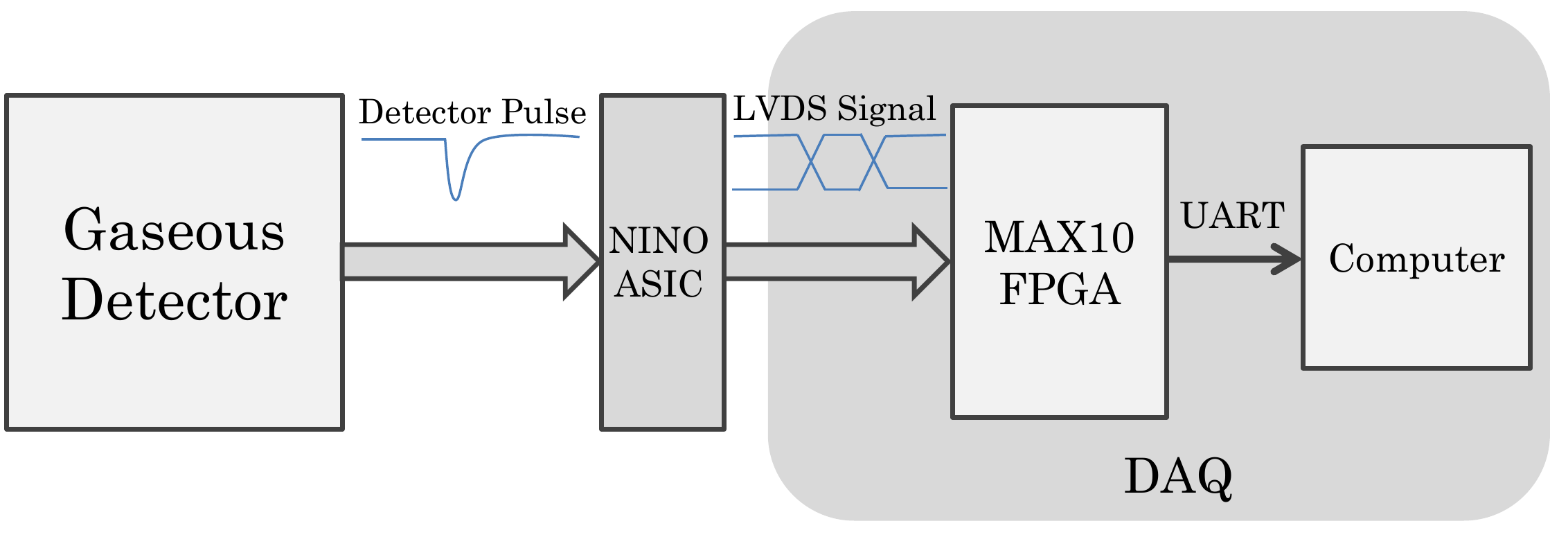}
\caption{Schematic diagram of the Data Acquisition System (DAQ)}
\label{schamatic DAQ}
\end{figure}

The tracking setup consists of some important components. A complete schematic diagram of the DAQ has been shown in figure~\ref{schamatic DAQ}. NINO ASICs are used as front-end electronics and FPGA in the back-end for storing the data. VHDL and Python language are used to program FPGA and for acquiring the data respectively. These components have been briefly described below.

\subsection{Front-end Electronics}
\label{sec:Front-end Electronics}
NINO ASICs are used as readout electronics, which is an ultra-fast front-end preamplifier discriminator chip that has been developed for use in the ALICE TOF detector. This ASIC chip has eight channels, each channel accepts signals from the detector and provides LVDS (Low Voltage Differential Signal) output. The output pulse width is dependent on the input signal charge and threshold value of NINO. Using this NINO ASIC, INO Collaboration developed a board that has been used for muon tracking experiments. This board also has eight channels and an adjustable threshold regulator.

\begin{figure}[h!]
\centering
\includegraphics[width=0.9\textwidth]{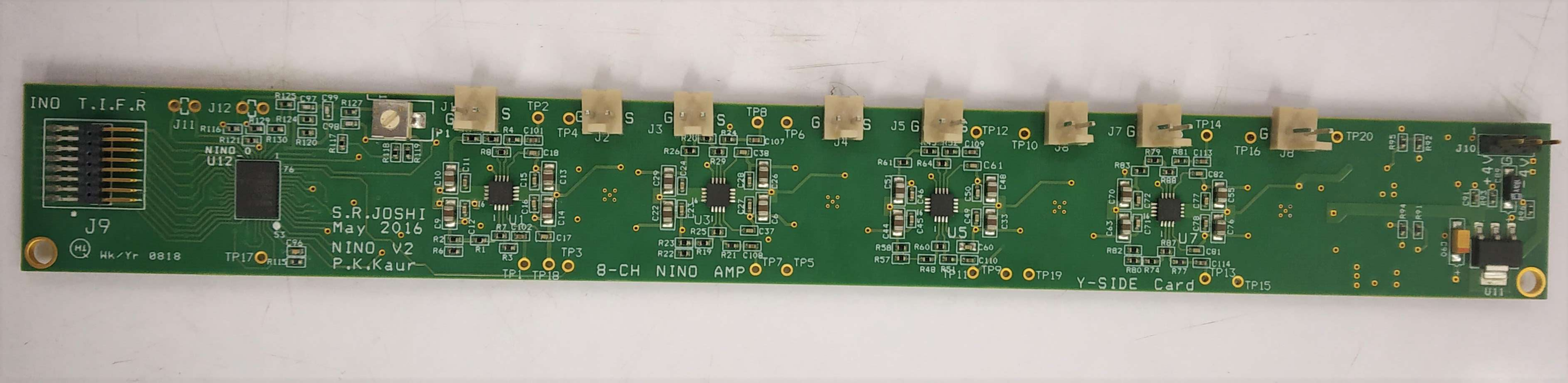}
\caption{NINO Board}
\label{fig:nino}
\end{figure}

\subsection{Back-end Electronics}
\label{sec:Back-end Electronics}
For such front-end a suitable back-end and Acquisition is required that can handle fast incoming LVDS signals from the NINO. For such a system FAGA is very useful due to its large number of I/O, availability, and cost-effectiveness. FPGA is a fabricated silicon device that can be electrically programmed to become almost any kind of digital circuit or system. FPGA can be considered as a simple logic device or a complete application. It contains thousands of logic elements that can be wired in different configurations according to the logical requirements of the application.

\begin{figure}[h!]
\centering
\includegraphics[height=0.3\textwidth]{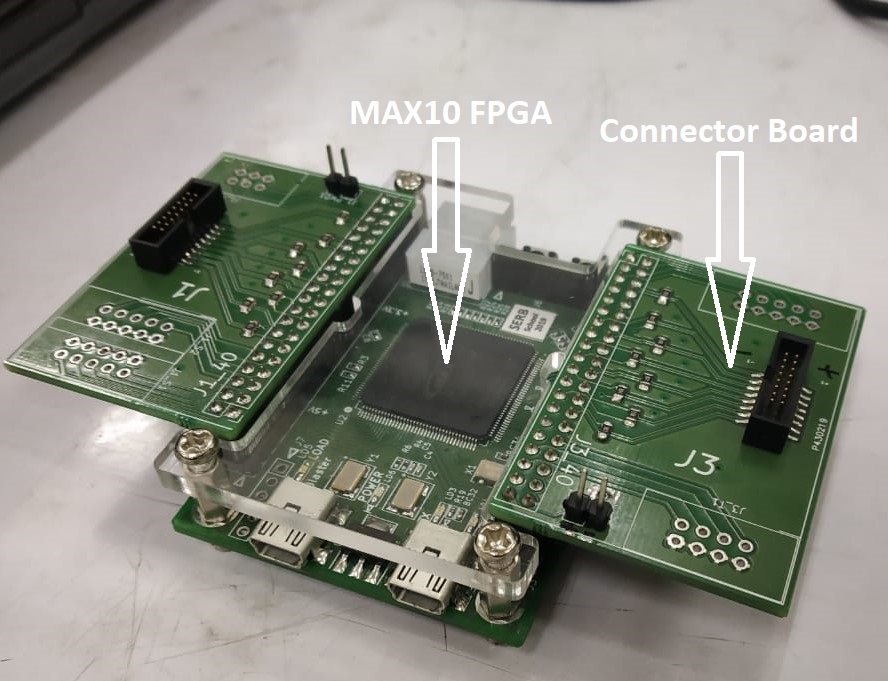}
\caption{FPGA board with connector}
\label{fig:expt setup}
\end{figure}

Intel MAX10 FPGA has been used for this purpose due to its capability of accepting LVDS signals as input and high bandwidth. The MAX 10 development board is a compact FPGA development platform developed by TIFR. The board contains Intel MAX 10 FPGA with 2000 Logic Elements (LEs), 108 memory blocks, 101 I/O pins, and a 50 MHz onboard clock. A connector board has been designed to map LVDS connections between NINO and FPGA board with the required 100~$\Omega$ termination resistance.

\subsection{Software Implementation}
\label{sec:Software Implementation}
\begin{figure}[h!]
\centering
\includegraphics[width=0.7\textwidth]{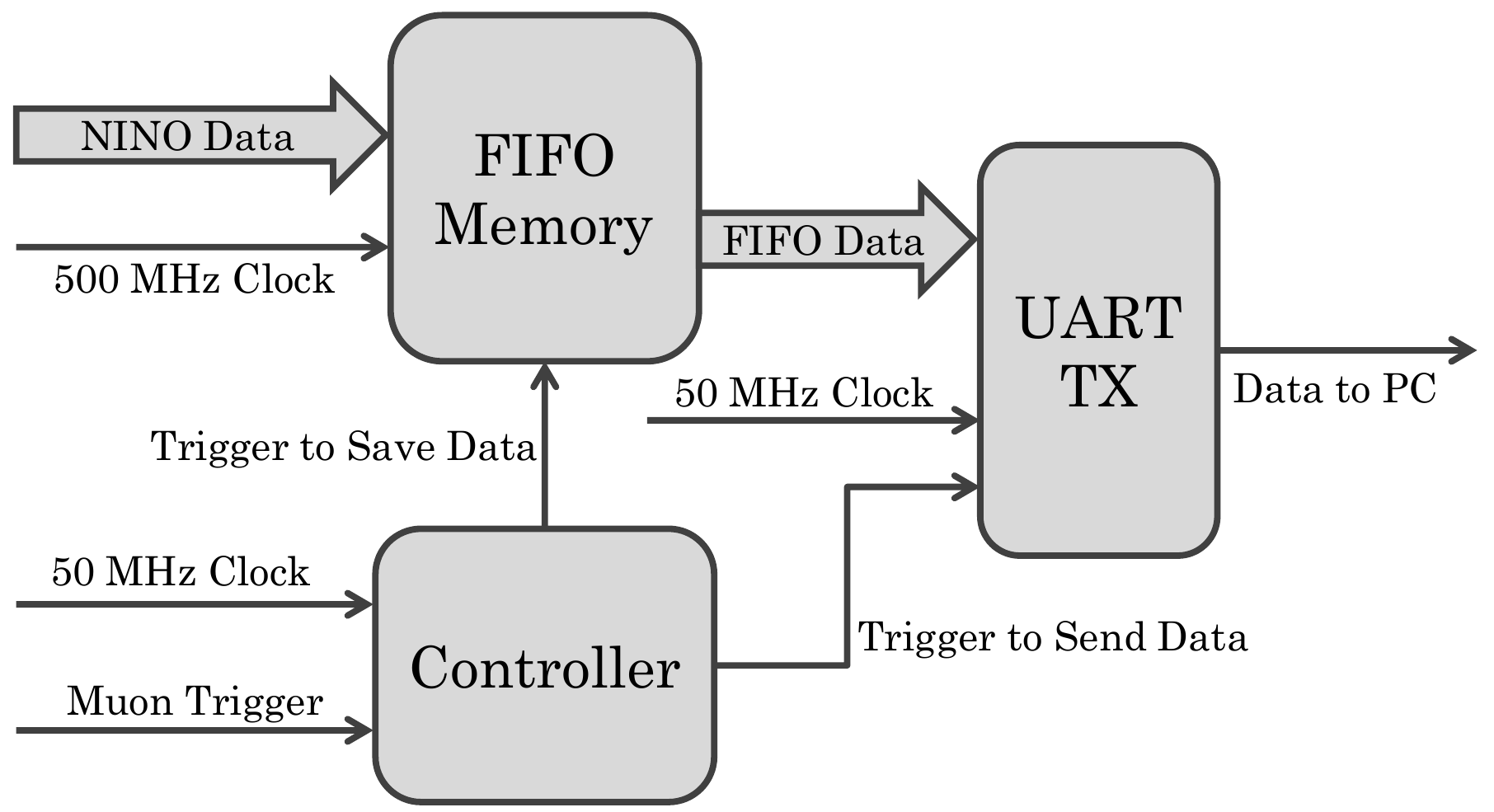}
\caption{Schematic diagram of the operation inside the FPGA}
\label{fig:FPGA dataflow}
\end{figure}

As discussed above FPGA is programmable hardware so the programming of FPGA is an important task for this experiment. VHDL and Verilog are two hardware description languages that are used to program FPGA. For this purpose, VHDL is used as a programming language and Quartus prime software has been used to compile and upload code to FPGA. UART protocol has been used to transfer data from FPGA to the computer, which has been implemented on FPGA. To measure the TOT output of NINO, a 500 MHz clock has been generated using a Phase-Locked Loop (PLL). This clock helps to measure TOT with a maximum probable inaccuracy of 2 ns. FPGA is programmed similar to an oscilloscope, a temporary FIFO memory is used to store NINO input. A controller, which sends the trigger signal for each nuon event to transmit the NINO data through the UART module which stands for Universal Asynchronous Receiver/Transmitter. This data flow algorithm has been implemented for all input channels inside the FPGA. The schematic workflow of the FPGA has been illustrated in figure~\ref{fig:FPGA dataflow}

Python programming language has been used to acquire data on PC using COM port and also for analysis and extracting position information of the muon track. The data contains strip hits and the pulse width of each channel of each NINOs. Input strips orientation is required to map the exact hit position.

\section{Tests of the Read-out System}
\label{sec:Tests of the Read-out System}
To verify the tracking capability, an experimental setup has been designed for this readout system. One RPC detector, one FPGA Development Board, and two NINO boards have been used for this test setup. The position of incoming cosmic muon has been detected by using a test area of 8X8 strips of RPC detector. The detector configuration and readout strips connections are shown in figure and figure respectively.

\begin{figure}[h!]
\centering
\includegraphics[height=0.30\textwidth]{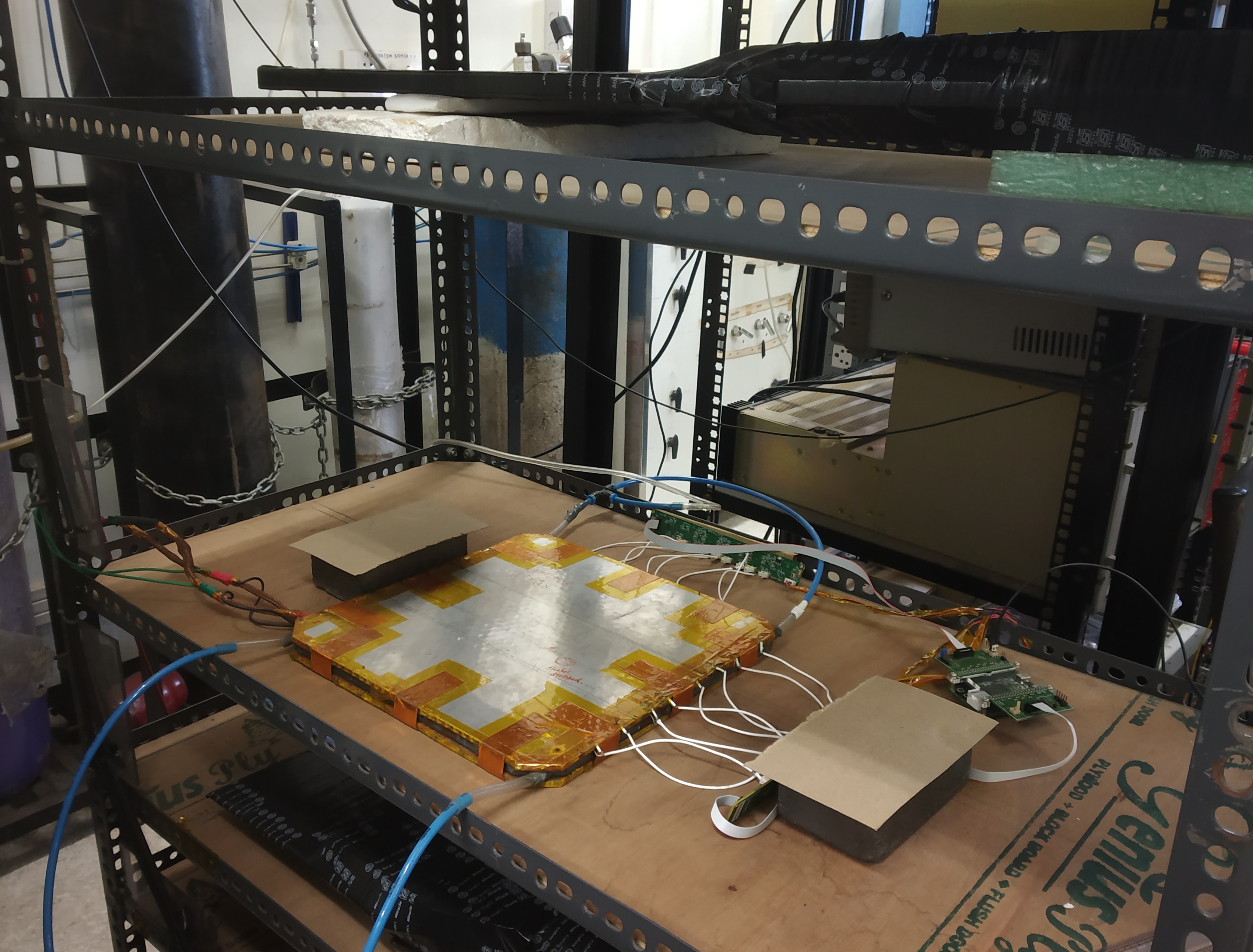}
\hspace{5mm}
\includegraphics[height=0.30\textwidth]{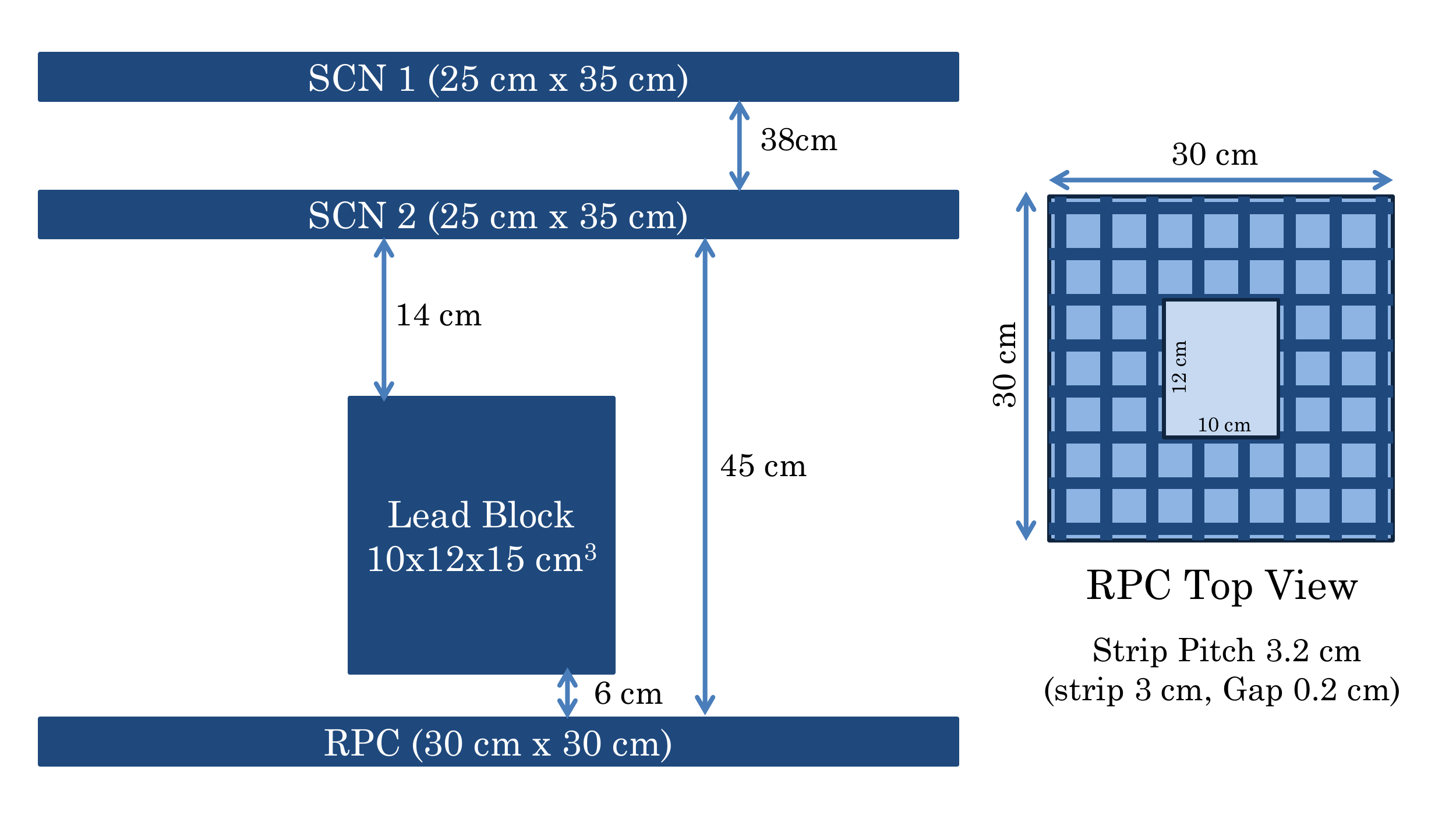}
\caption{Image of the experimental setup}
\label{fig:expt setup}
\end{figure}

\subsection{Description of Test Setup}
\label{sec:Description of Test Setup}
In our test setup, a glass-RPC of dimension 30cm x 30cm, filled with a gas mixture of 95\% Freon and 5\% Isobutane, equipped with two orthogonal panels of readout strips of width 3 cm and pitch 3.2 cm, has been operated. An 8-channel ultra-fast front-end preamplifier discriminator chip NINO has been used to obtain Time-Over-Threshold (TOT) outputs corresponding to the current signals induced on the readout strips due to a muon event. For each readout plane, 1 NINO board has been connected to acquire signals from 8 strips. Total 16 strips and two NINOs have been used to cover 8 $\times$ 8 strips configuration. Each NINO chip provides as outputs 8 pairs of LVDS (Low Voltage Differential Signalling) signals which are suitable for low-power, high-speed transmission with better noise immunity. For parallel post-processing and to take advantage of this LVDS scheme, we have used an FPGA development board with ALTERA MAX 10, which has dedicated pins for the direct acquisition of LVDS analog signals. The FPGA is programmed to transfer the acquired data to the computer using the UART protocol. The RPC data for muon events as collected by the present readout scheme have been analyzed to estimate the position information. The RPC has been placed in a horizontal plane to maximize cosmic ray interaction. Two plastic scintillators with an overlapped region that covers nearly 25 cm $\times$ 35 cm of the readout strips of the RPC used as muon trigger. A lead block of dimension 12 cm $\times$ 10 cm $\times$ 15 cm has been placed 6 cm above the middle of RPC. The image of the experimental setup has been shown in figure~\ref{fig:expt setup}.

\subsection{Test Results}
\label{sec:Test Results}

\begin{figure}[h!]
\centering
\includegraphics[height=0.42\textwidth]{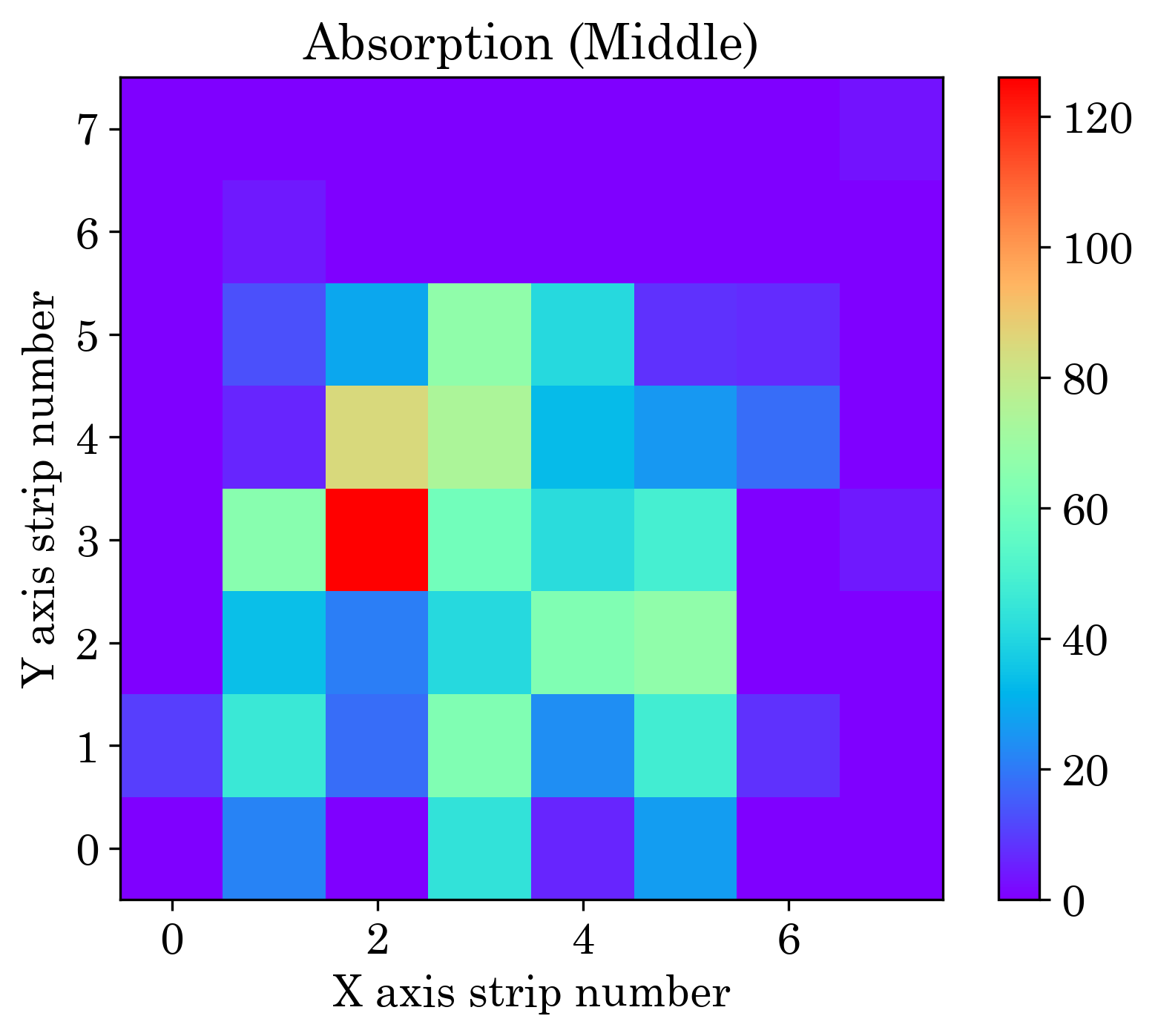}
\hspace{5mm}
\includegraphics[height=0.42\textwidth]{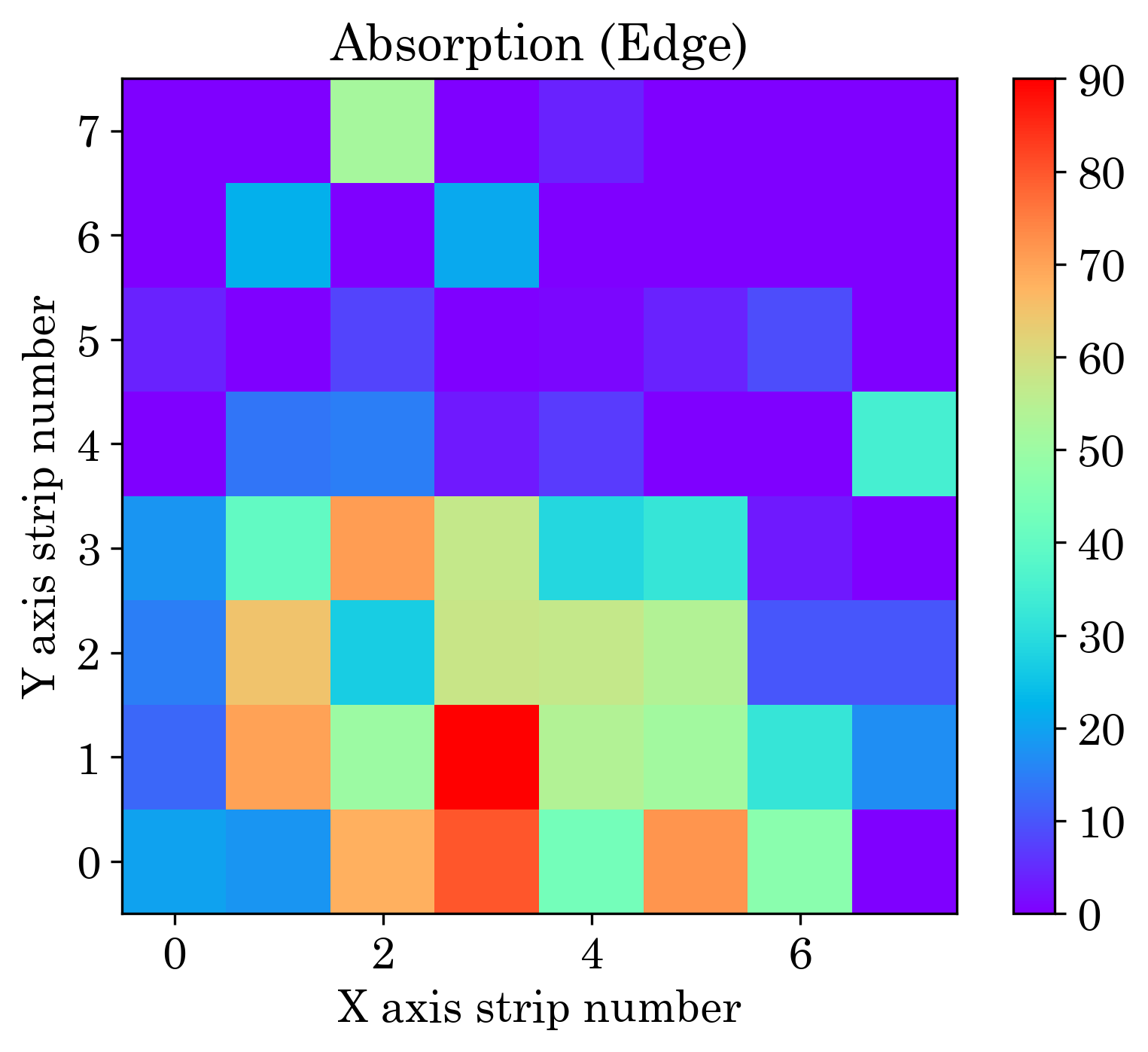}
\caption{Experimental Muon Absorption}
\label{fig:expt absorption}
\end{figure}

In our tracking experiment, multiple numbers of event data from two NINOs have been used to get hit information along X and Y direction, so from this data particle position in the 2D plane of dimension, 8x8 can be extracted. The weighted mean position has been used for the multiple hit events. The valid events have been filtered out using the strip hit and pulse width information. For simplicity, Only avalanche events have been used for muon tracking. The event with greater than four strip hits has been treated as a streamer event and the event single-sided hit has been excluded from muon track. The hit points have been put into 8 $\times$ 8 pixels for both background hits and hits using lead block in line of muon tracks. The absorbed muon has been calculated by subtracting absorbed data from background hits. Similar studies have been carried out using the different positions of lead block middle and edge of the RPC. The experimental data for muon absorption pattern for two different positions have been shown in figure~\ref{fig:expt absorption}.

\section{Comparison with Geant4 Simulation}
\label{sec:Readout Scheme}
In order to verify our experimental absorbed data by the lead block, a simulation study has been carried out by Geant4~\cite{07} library, which is a well-known simulation package for high energy and particle physics simulation. 

A model, similar to our experimental setup containing one position-sensitive gaseous ionization detector for detecting the muons has been constructed in Geant4. The setup has been subjected to a limited exposure of the cosmic muon flux received at the sea level using Cosmic Ray Library (CRY)~\cite{08}. The hits obtained from the detector have been used to get the absorption pattern.

FTFP\_BERT physics list has been used to propagate the particles throughout the lead and detector. It takes care of all the electromagnetic and weak interactions of the charged particles and gamma radiations. The hit information from the detectors has then been extracted from the muon track.

\subsection{Simulation Result }
\label{sec:Front-end Electronics}

\begin{figure}[h!]
\centering
\includegraphics[height=0.42\textwidth]{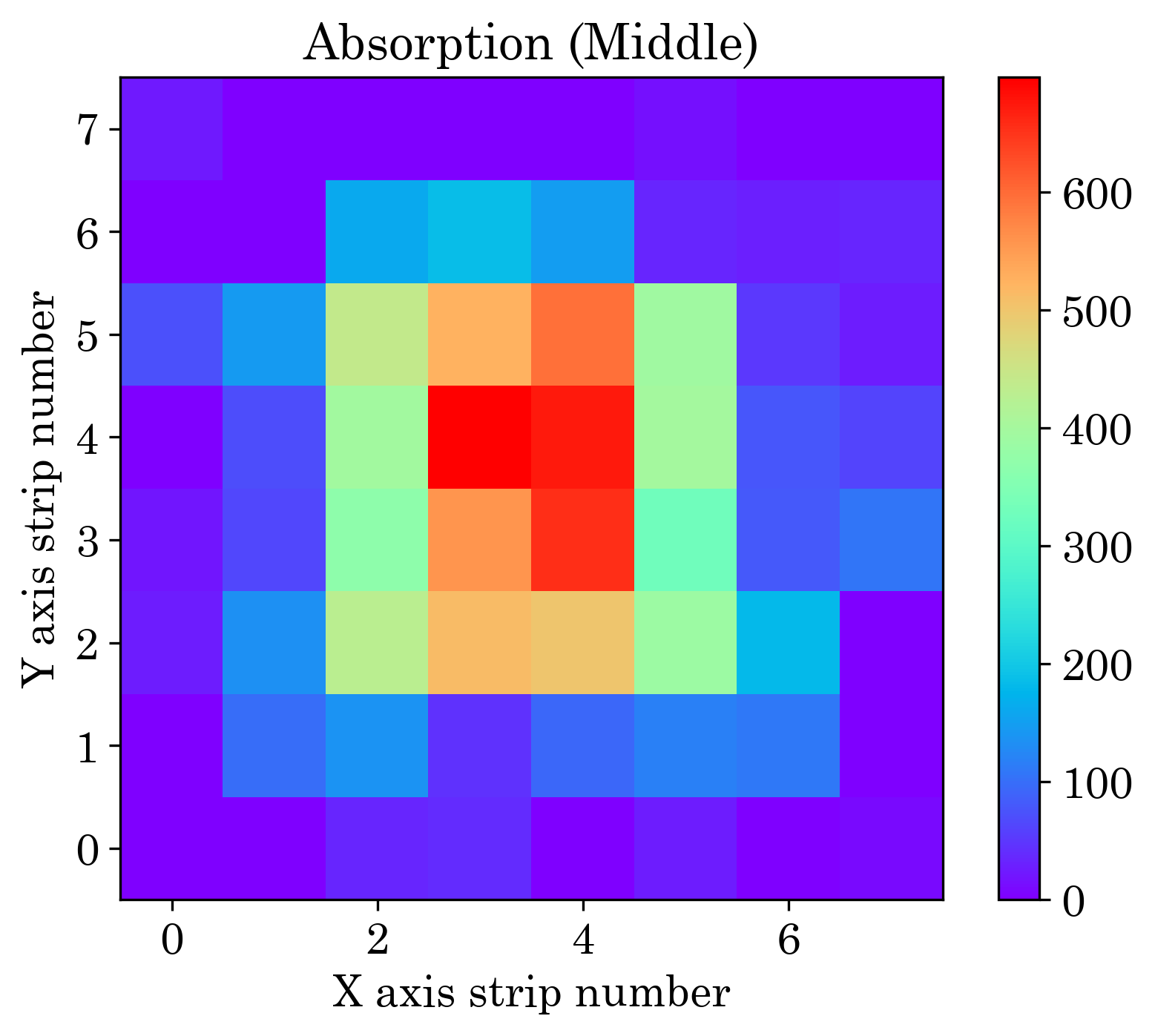}
\hspace{5mm}
\includegraphics[height=0.42\textwidth]{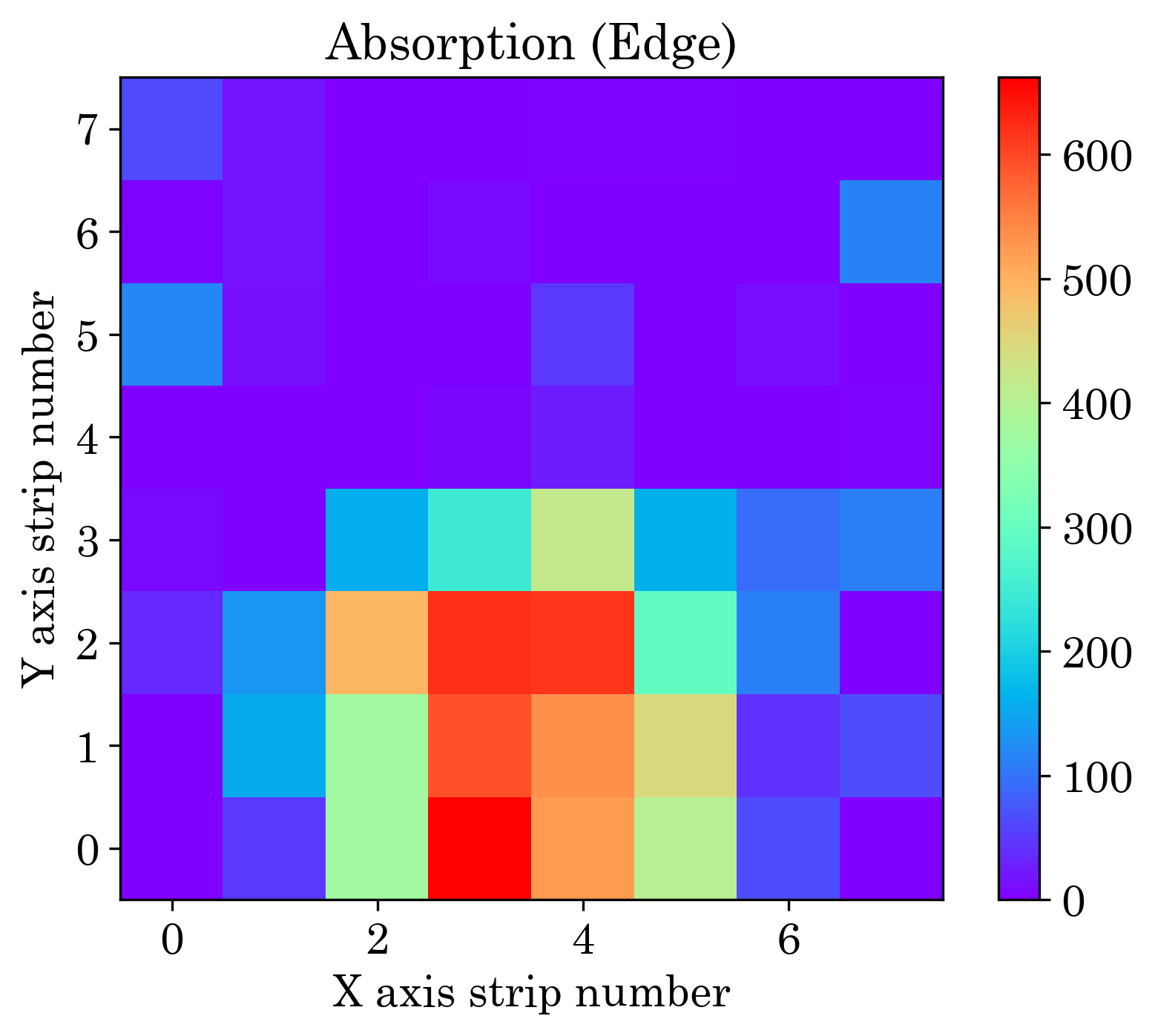}
\caption{Simulated Muon Absorption}
\label{fig:sim absorption}
\end{figure}

The images produced by the setup have been described above, The images have been segmented into 8 $\times$ 8 pixels, each having an area of 3.2 $\times$ 3.2 $cm^2$. Similar to the experimental method every hit point has been put into 8 $\times$ 8 pixels for both cases. The absorbed muon has been calculated by subtracting absorbed data from background hits for two different positions of the lead block. The simulated data for muon absorption pattern for two different positions have been shown in figure~\ref{fig:sim absorption}.

\section{Conclusions}
\label{sec:Conclusions}
The FPGA-based DAQ prototype has been tested with NINO and RPC detector using cosmic muon. The position of muon hit has been detected by using this DAQ. The experimental results are similar to the simulated result. To get a better position resolution, the authors have the plan to use a readout with a thinner width and separation gap.

\section*{Acknowledgements}
The authors are thankful to Mr. Shaibal Saha and other members of our laboratory for the experimental support and advice. The authors are extremely grateful to the INO Collaboration for the support in electronics also thankful to ANP Division, SINP, UGC and Govt. of India.

\bibliographystyle{unsrt}

\begin{thebibliography}{99}

\bibitem{01} S. Tripathy {\it et al.}, JINST {\bf 15},  C11013 (2020)
\bibitem{02} G. Jonkmans {\it et al.}, Ann. Nucl. Energy {\bf 53}, 267 (2013).
\bibitem{03} K.N. Borozdin {\it et al.}, Nature {\bf 442}, 277 (2003).
\bibitem{04} H.A. Bethe, Phys. Rev. {\bf 89}, 1256 (1953).
\bibitem{05} V.L. Highland, Nucl. Instrum. Meth. {\bf 129}, 497 (1975).
\bibitem{06} G.R. Lynch and O.I. Dahl, Nucl. Instrum. Meth. B {\bf 58}, 6 (1991).
\bibitem{07} GEANT4 collaboration, Nucl. Instrum. Meth. A {\bf 506}, 250 (2003).
\bibitem{08} C. Hagmann {\it et al.}, IEEE Nuclear Science Symposium Conference Record, 1143–1146 (2007).

\end{thebibliography}

\end{document}